# It's Time to Do Something: Mitigating the Negative Impacts of Computing Through a Change to the Peer Review Process

**Brent Hecht** (Northwestern University)**, Lauren Wilcox** (Georgia Tech)**, Jeffrey P. Bigham** (Carnegie Mellon University)**, Johannes Schöning** (University of Bremen)**, Ehsan Hoque** (University of Rochester)**, Jason Ernst** (RightMish, University of Guelph)**, Yonatan Bisk** (University of Washington)**, Luigi De Russis** (Politecnico di Torino)**, Lana Yarosh** (University of Minnesota)**, Bushra Anjam** (Amazon, Inc.)**, Danish Contractor** (IBM Research, IIT New Delhi)**, Cathy Wu** (University of California, Berkeley)

**INTRODUCTION**

The computing research community needs to work much harder to address the downsides of our innovations. Between the erosion of privacy [6,8,9,29,31], threats to democracy [16,27], and automation's effect on employment [2,4,12,35] (among many other issues), we can no longer simply assume that our research will have a net positive impact on the world.

While bending the arc of computing innovation towards societal benefit may at first seem intractable, we believe we can achieve substantial progress with a straightforward step: **making a small change to the peer review process**. As we explain below, we hypothesize that our recommended change will force computing researchers to more deeply consider the negative impacts of their work. We also expect that this change will incentivize research and policy that alleviates computing's negative impacts.

The current status quo in the computing community is to frame our research by extolling its anticipated benefits to society. In other words, **rose-colored glasses are the normal lenses through which we tend to view our work**. This Pollyannaish perspective is present in our research papers, our applications for funding (e.g. NSF proposals), and our industry press releases. For instance, computer scientists who seek to automate yet another component of a common job description point to the merits of eliminating so-called "time consuming" or demanding tasks. Similarly, those working on generative models praise the quality of the audio and video (e.g. "deepfakes") they generate, gig economy and crowdsourcing researchers highlight the reduced costs of crowd workflows, and so on. Indeed, in many sub-fields of computer science, it is rare to encounter a paper, proposal, or press release that does not use these types of pro-social framing devices.

However, one glance at the news these days reveals that focusing exclusively on the positive impacts of a new computing technology involves **considering only one side of a very important story**. Put simply, the negative impacts of our research are increasingly high-profile, pervasive, and damaging. Driverless vehicles and other types of automation may disrupt the careers of hundreds of millions of people [2,4,35,36]. Generated audio and video might threaten democracy [26,27]. Gig economy platforms have undermined local governments and use technology for "regulatory arbitrage" [10,20]. Crowdsourcing has been associated with (and sometimes predicated upon) sub-minimum wage pay [7,18]. And it doesn't stop there. Mobile devices have increased vehicular deaths by the thousands [19,23]. The technologies we have designed have contributed to a sudden and rapid decline in privacy rights [6,8,9,29,31]. Companion robots are altering the meaning of relationships [33]. Persuasive technology design is eroding our attention and may have addictive qualities [3]. Social media platforms are facilitating the spread of false information, conspiracy theories, and propaganda [30]. The list goes on and on.

There clearly is a massive gap between the real-world impacts of computing research and the positivity with which we in the computing community tend to view our work. We believe that this gap represents **a serious and embarrassing intellectual lapse.** The scale of this lapse is truly tremendous: it is analogous to the medical community only writing about the benefits of a given treatment and completely ignoring the side effects, no matter how serious they are.

What's more, **the public has definitely caught on to our community-wide blind spot and is understandably suspicious of it**. Consider this recent tweet thread from the comedian Kumail Nanjiani of *Silicon Valley* fame. Nanjiani writes with exasperation that when he visits tech companies and conventions, he sees "scary" technologies whose creators have clearly given their negative implications "ZERO consideration." Nanjiani reports that when he asks about these negative implications, our fellow members of the computing community, "don't even have a pat rehearsed answer." A similar dynamic can be observed in many of our community's other interactions with the public, e.g., a recent RadioLab interview about research on generative visual and audio models and their powerful anti-social applications [26].

At our inaugural ACM Future of Computing Academy meeting last June, many of us agreed that the computing research community must do more to address the downsides of our innovations. Indeed, our view is that it is our moral imperative to do so. After several months of discussion, an idea for acting on this imperative began to emerge: we can **leverage the gatekeeping functionality of the peer review process**.

Below, we describe our specific recommended change to the peer review process. If widely adopted, we believe that this recommendation can make meaningful progress towards a style of computing innovation that is a more unambiguously positive force in the world. We expect that a large proportion of our readers are peer reviewers themselves. If you are a peer reviewer, in many communities, you may be able to **try out our recommendation immediately, applying it to the next paper that appears on your review stack and citing this post as justification**. In other communities, implementing our recommendation will have to be part of a larger discussion within your community. We hope that you will tell us about these discussions and your experiences implementing our recommendation. This post is part of the FCA Discussions series, which means our recommendations are intended to start a conversation rather than end one!

**SMALL CHANGE TO PEER REVIEW, BIG IMPACT?**
At a high level, our recommended change to the peer review process in computing is straightforward:

> *Peer reviewers should require that papers and proposals rigorously consider all reasonable broader impacts, both positive and negative.*

Papers and proposals, as noted above, are typically already flush with anticipated positive impacts. As such, this recommendation will in practice involve insisting that authors consider potential negative impacts.

For example, consider a grant proposal that seeks to automate a task that is common in job descriptions. Under our recommendation, reviewers would require that this proposal discuss the effect on people who hold these jobs. Along the same lines, papers that advance generative models would be required to discuss the potential deleterious effects to democratic discourse [26,27] and privacy [28].

Importantly, reviewers should not only require that potential positive and negative impacts simply be mentioned, but also that they be **strongly motivated**. Hand-waving is too often permitted by reviewers even when discussing positive impacts, and this understandably decreases public trust in otherwise highly-rigorous research. A motivation of sufficient strength will likely come through references to prior work both inside and outside computing, as well as to articles in reputable media outlets.

The specific mechanisms of the discussion of impacts in research papers will likely vary from subfield to subfield. However, one option that is tractable across much of computing is for authors to add a "Broader Impacts" or a "Societal Impacts" section near the end of a paper, a la "Future Work" and "Limitations". This section would **summarize both the anticipated positive and negative impacts of the paper and motivate these anticipated impacts with the proper citations**. This section may also make an explicit argument that the paper's contributions will have a net positive impact. For grant proposals, the appropriate venue for this discussion will likely be more obvious. For instance, the U.S. National Science Foundation already requires that authors include a "Broader Impacts" section. In these cases, our recommendation is simply that reviewers insist that this section include discussion of negative broader impacts as well as positive broader impacts.

Our recommendation states that authors should engage with all "reasonable" positive and negative impacts. **This raises a critical question: What is "reasonable"?** As discussed below, we suggest that peer reviewers initially adopt a "big tent", author-friendly approach. In this vein, initial thresholds for "reasonable" might emerge from the nature of our recommendation as a means of making computing research more accountable to the public. For instance, one initial big tent threshold might be the following: if there has been significant media coverage of a negative impact that is relevant to a paper or proposal, that impact should be addressed by the authors of that paper or proposal.

Moving forward, one could imagine the community developing more formal reviewer guidelines, e.g. guidelines based on ACM's soon-to-be-released revised code of ethics or other relevant frameworks. We also expect that some sub-fields will want to have different formal guidelines than other sub-fields.

**What if a research project will do more harm than good?**
If successful, our recommendation will make the expected positive and negative impacts of our community's research much more transparent. We expect that doing so will highlight a deeply uncomfortable truth: some research in computing does more harm than good, at least when considered in isolation. This then raises a critical question: what should be done with a paper or proposal whose impacts are clearly net negative?

In the case of a paper or proposal with a likely net negative impact, we first recommend that authors be encouraged to **discuss complementary technologies, policy, or other interventions that could mitigate the negative broader impacts** of the paper or proposal. In other words, authors should outline future work for researchers, practitioners, and policymakers that can help ensure that the contribution of the paper or proposal will have a net positive impact. In most cases, we expect that this approach will help researchers outline a path by which their work can have an overall positive effect. This approach will also have additional benefits, in particular **increased formal and public support from the computing community for policy that can mitigate computing's negative impact**. Indeed, we recommend all papers and proposals consider listing complementary advances in technology and policy that can mitigate potential negative effects, even if the paper or proposal has a clear argument that its net effect will be positive absent these advances.

But what about papers and grant proposals that cannot generate a reasonable argument for a net positive impact even when future research and policy is considered? For grant proposals in this situation, the implications are straightforward: **it is unlikely that government agencies will want to use taxpayer dollars for research that is, on balance, going to hurt the taxpayers who paid for the grant**. No matter how intellectually interesting an idea, computing researchers are by no means entitled to public money to explore the idea if that idea is not in the public interest. As such, we recommend that reviewers be very critical of proposals whose net impact is likely to be negative.

Research papers that are likely to have a net negative impact present a trickier problem than grant proposals in the same situation. Should ACM KDD, AAAI, NIPS, ACM SIGCHI, etc. publish papers that reviewers and authors agree will do more harm than good, even when considering potential future research and policy? We believe the answer is 'yes', but with an important caveat: **the research community should more actively evaluate its members on their ethical decision-making when it comes to research project selection**. In doing so, we would simply be extending a well-established practice when it comes to ethics in the research process. If one has plagiarized or violated other ethical standards in the research process, it is appropriately difficult to acquire or keep a job, tenure or a promotion. The same should be true for someone who continuously chooses to conduct anti-social research, no matter how rigorous that research.

**What about non-scholarly research? (The role of the tech press)**

We are fortunate in computing that our research activity has heterogeneous outputs. While much of our research appears in peer-reviewed scholarly publications, much of our research also manifests in products or services. While research that does not go through the peer review process tends to have more limited gatekeeping infrastructure, **there is a set of powerful actors that can play a similar role to that which we recommend for peer reviewers: the technology press**. The tech press frequently also acts as a secondary gatekeeper for scholarly publications.

As such, **we encourage the press to hold accountable all public communication regarding computing innovation in the same fashion as we suggest for peer reviewers above.** This means asking researchers and the firms that represent them to enumerate the downsides of their innovations. It also means asking them to discuss what changes to their technologies and what new policy might mitigate these downsides. If a researcher or firm can only highlight positive potential impacts, the press should potentially contact other researchers to get a sense of potential negative impacts. A simple heuristic might be to approach all stories about the innovations from our community with the following lens: "Is the impact of this technology likely to be a net positive or a net negative for society?"

We note that adopting this recommendation is particularly important for covering research that is published in pre-publication portals like ArXiv, in which claims towards impact (as well as all other claims) are not vetted by peer reviewers. Some of us would suggest eschewing the coverage of research in pre-publication portals more generally.

It is also important to note that in many cases, **the tech press is way ahead of the computing research community on this issue**. Tech stories of late frequently already adopt the framing that we suggest above.

**EXPECTED OUTCOMES**

We expect that action on the above recommendations will lead to a number of desirable outcomes:

- **Increased intellectual rigor:** Our scholarly work will have increased intellectual rigor in that statements about the impacts of our work will no longer consider only desirable evidence.

- **Technology with greater societal benefit:** Researchers will be incentivized to change the technologies they create to tilt the scales towards more positive outcomes.

- **More support for key governmental policies:** Policy that mitigates the downsides of existing and new technologies will receive much more public support from the computing community.

- **Larger incentives for research that mitigates negative impacts:** Research that mitigates the downsides of existing and new technologies will be more strongly incentivized and motivated.

- **More engagement with the social sciences:** We will be encouraged to engage more with the social science literature, which has a great deal of expertise in understanding the societal impacts of diverse types of interventions (as well as an understanding of the limitations of our ability to know these impacts).

More generally, the computing research community will more deeply consider the negative implications of our work, something that is currently a major intellectual oversight, something that we believe is critically important, and something that the public is increasingly demanding.

Of course, we also anticipate a number of questions and challenges associated with our recommendations:

First and foremost, we expect that authors may be concerned that it is impossible to predict with high certainty the universe of use cases for a technology. Given this uncertainty, authors may believe they should not be expected to discuss anticipated positive and negative impacts. **There is no doubt that the above proposal will involve a great deal of reasoning in the face of uncertainty** by both authors and reviewers. However, we believe that the cost of that uncertainty is reasonable to bear for several reasons. Primarily, we are already bearing this cost, it is just implicit. A large percentage of our work is motivated either directly or indirectly by claims about positive impacts. For instance, many large grant proposals make explicit arguments about the good they will do for the world. We are simply suggesting that these claims be made in a maximally rigorous way by deeply considering the potential negative impacts as well. We also believe a significant portion of use cases for a given innovation can be predicted with reasonable certainty, even if doing so is undesirable. For instance, that autonomous vehicles could severely disrupt the job market was obvious many years ago. The same is true of many of the more negative use cases for social media, GANs, the sharing economy, and so on. Put simply, if Kumail Nanjiani can do this, so can we.

Second, **we expect there may be some disagreements between authors and reviewers** about what constitutes a negative impact and how to compare "positive impact apples" to "negative impact oranges". However, in many cases, our recommendations do not require author-reviewer agreement on the valence of a net or gross impact: just that impacts be listed. In the case of grant proposals in which such a judgment is necessary, we recommend that authors and reviewers defer to a big tent interpretation of our recommendations, at least until other standards are developed.

Third, we expect that **it will take a non-trivial amount of time for norms and standards to develop** in the field of computing and its various sub-fields. During that time, some papers and proposals may bear an unfair share of the transition costs, even considering our suggestion to initially defer to big tent interpretations of our recommendation. This is particularly the case for interpretations of "reasonable" in our recommendation above. We believe, however, that the costs of this unfairness are minimal compared to the benefits of implementing our recommendation, and we fully expect that some of our papers and proposals may be the ones to bear the cost!

Fourth, we expect **that some authors and reviewers may not take their duty to engage with broader impacts as seriously as others**. In these cases, "boilerplate" statements may emerge that are used for many papers in specific research areas. We discourage the use of these statements as they risk diminishing the level of rigor present in any paper that is published in a significant computing venue. However, even such boilerplate statements will be an improvement upon the status quo, in which papers almost never consider negative impacts at all.

## A FEW EXAMPLES

To make the above recommendations more concrete, we provide below a few examples of how researchers might adapt to the peer review change that we suggest above.

**Crowdwork**: A researcher who invents a new crowdwork framework likely motivates her work by highlighting the problem the framework solves and often the financial benefits of the solution. Crowdwork, however, also comes with serious negative externalities such as incentivizing very low pay [18]. Under our recommendations, this researcher should ideally find ways to engineer her crowdwork framework such that these externalities are structurally mitigated. Alternatively or additionally, she should state what new technologies or policy must complement her system for it to have clear net positive impact. For instance, she might advocate for minimum wage laws to be adapted to a contingent labor context. She should also ensure that her system still has practical use in the context of higher pay.

**Blockchain**: Consider a researcher writing a manuscript or whitepaper describing a new blockchain-based technology. Under current practices, this researcher or practitioner would almost certainly not address the serious negative externality of blockchain's energy usage and corresponding carbon footprint [25]. Under our recommendation, this researcher would be urged by gatekeepers (peer reviewers or the press) to discuss this significant, often-unspoken downside to blockchain-based approaches. This is especially the case if – like many blockchain-based technologies – some of the key short-term use cases for the new blockchain-based technology are to better support some societal function in developing countries. Since many of these countries are expected to bear very heavy costs from climate change [34], that the new technology contributes to climate change may significantly complicate the claimed benefits of the technology.

**Social Media**: There's no need for a hypothetical example here. David Ginsberg and Moira Burke – researchers at Facebook – have provided a roadmap for how to implement some of these recommendations within an industry context with their blog post titled "Hard Questions: Is Spending Time on Social Media Bad for Us?" [11] In their post, they discuss how their research (as manifest in the Facebook product) contributes to positive and negative well-being, as well as what Facebook is doing to mitigate the negative effects. The technology press reacted with astonishment to the consideration of negative and net impacts by a technology company, calling the post "quietly groundbreaking" [21]. Our view is that this should not be astonishing; it should be a norm.

**Accessible Technology:** Consider a researcher who develops a new access technology that supports people with disabilities in independently doing something that used to require human help. This technology may dramatically decrease the cost of providing this support and make the service more easily available to people who need it. However, if the services were previously provided by an employee, the new technology may lead to fewer jobs. Also, this technology likely does not have a social component, which users of the service may have highly valued, but which is now unavailable to them. The researcher who develops this technology would be required under our recommendations to enumerate not only the benefits to accessibility, but also the negative impacts on employment and social interaction. The researcher would also likely choose to highlight policies or new technologies that could mitigate these negative impacts (e.g. the integration of remote social support technologies [22,38]).

**Mobile and Digital Health:** In public health, inequalities can result not only from differences in access to technologies, but also from differences in how technology – and the data collected through it – is used (and by whom). Consider a hypothetical academic research paper submitted to a computing conference that applies data science to behavioral or social media data to develop predictive models of illness based on behavioral signatures or linguistic cues [37]. The authors frame the contributions as positive, and they tout the benefits of earlier diagnoses and earlier intervention for health concerns. They might wish to demonstrate how variance in movement patterns or social media use suggest opportunities for health interventions. Under current norms, the paper might address some limitations of the scientific approach, but would almost certainly fail to discuss the implications of tracking and sharing the personal data of large numbers of individuals [32]. These implications suggest a great number of broader negative impacts, including threats to national security and disparities in insurance coverage that in turn put millions of lives at risk and threaten our economy [24,32]. Of course, the list of foreseeable concerns does not end there. By prioritizing the collection and measurement of data to those who already have access to Internet resources and personal computing devices, we also skew the composition of the data set that is used to generate models of illness and health needs.

Under our peer review recommendations, many health papers – including our hypothetical example – would be rejected for inaccurately representing the broader impacts of the research contributions. To be successful, such a paper would need to acknowledge and discuss its potential impact on privacy, security, and discrimination in a rigorous fashion. It would also be advisable (although not required) for the paper to make a convincing argument for (a) the net positive impact of the contributions in light of these negative broader impacts, and (b) what additional research or policy would be needed for the paper to have a net positive impact.

**Automation-related Research**: Many national funding agency (e.g., NSF) proposals from the computing community involve the automation of tasks that appear in the job descriptions of entire sectors of workers. Proposals of this type tend to be particularly bad offenders when it comes to focusing only on the positive impacts of computing research and ignoring the negative impacts. For instance, consider a hypothetical proposal that seeks to advance robotics such that they can be used to automate home care for older adults and those with physical disabilities. This proposal would likely highlight how it will reduce the costs of home care and eliminate "repetitive" or "time-consuming" tasks for workers. However, under current norms, the proposal would also entirely omit discussion of the potential large-scale job loss that could occur if the corresponding research were to be successful (there are over one million people working in the home healthcare sector in the United States alone [5]). This is a particularly serious issue in the national funding agency context as it means that people are likely paying for research that may threaten their source of income with the taxes they pay on that income.

Under our peer review recommendations, many automation-related proposals – including our hypothetical example – would be rejected for inaccurately representing the broader impacts of the proposed research. To be successful, such a proposal would need to acknowledge and discuss its potential impact on job loss in a rigorous fashion. It would then need to make a convincing argument for (a) the net positive impact of the proposal itself or (b) what additional work would be needed for the proposal to have net positive impact.

**Storage and Computation:** Recent advances in storage systems and Graphical Processing Unit (GPU) processing afford the easy storage of massive amounts of data and the real-time computation on these data. This has incentivized corporations to collect every possible data point about their users, save this data indefinitely, and strive to monetize this data in new ways. While allowing for impressive new capabilities [1], this trend also presents tremendous risks to privacy [13,15,17]. Under our recommendations, researchers working in storage and GPU processing should consider these and other foreseeable potential risks in their papers. They should also enumerate technological and policy means by which these risks might be mitigated (e.g. technologies to automate General Data Protection Regulation (GDPR) required capabilities [14] and improvements to GDPR-like policies).

## WRAPPING THINGS UP

In this post, we have outlined a change to the peer review process in computing that could help the computing community take more responsibility for the negative implications of the technologies that we create. For those of you who are peer reviewers yourselves, we encourage you to, when possible, adopt our recommendations for papers in your review queue. If you do so, we would love to hear about your experiences. We also encourage feedback from the broader computing community more generally! Please feel free to contact any of us below or tweet at us @ACM_FCA.

*This blog post is part of the ACM Future of Computing's "FCA Discussions" series. Posts in this series are intended to spur discussion and do not consist of final, formal recommendations. The FCA is a new organization and our intention is for the discussions that emerge from these posts to inform the actions we ultimately take to address the underlying issues.*

## ACKNOWLEDGEMENTS

The authors would like to thank Moshe Vardi for his contributions to our discussions.

## REFERENCES

1. Michael Armbrust, Armando Fox, Rean Griffith, Anthony D. Joseph, Randy Katz, Andy Konwinski, Gunho Lee, David Patterson, Ariel Rabkin, Ion Stoica, and Matei Zaharia. 2010. A View of Cloud Computing. *Commun. ACM* 53, 4: 50–58. https://doi.org/10.1145/1721654.1721672

2. Anita Balakrishnan. 2017. Goldman Sachs analysis of autonomous vehicle job loss. *CNBC*. Retrieved March 20, 2018 from https://www.cnbc.com/2017/05/22/goldman-sachs-analysis-of-autonomous-vehicle-job-loss.html

3. Bianca Bosker. 2016. The Binge Breaker. *The Atlantic*. Retrieved March 26, 2018 from https://www.theatlantic.com/magazine/archive/2016/11/the-binge-breaker/501122/

4. Erik Brynjolfsson and Andrew McAfee. 2016. *The Second Machine Age: Work, Progress, and Prosperity in a Time of Brilliant Technologies*. W. W. Norton & Company.

5. Bureau of Labor Statistics. 2018. *Home Health Aides and Personal Care Aides*. United States Department of Labor, Washington, DC, USA. Retrieved from https://www.bls.gov/ooh/healthcare/home-health-aides-and-personal-care-aides.htm

6. Andrew Burt and Dan Geer. 2017. Opinion | The End of Privacy. *The New York Times*. Retrieved March 20, 2018 from https://www.nytimes.com/2017/10/05/opinion/privacy-rights-security-breaches.html

7. Michelle Chen. 2015. Is Crowdsourcing Bad for Workers? *The Nation*. Retrieved March 25, 2018 from https://www.thenation.com/article/crowdsourcing-bad-workers/

8. Martin Enserink and Gilbert Chin. 2015. The end of privacy. *Science* 347, 6221: 490–491. https://doi.org/10.1126/science.347.6221.490

9. Gregory Ferenstein. 2013. Google's Cerf Says "Privacy May Be An Anomaly". Historically, He's Right. *TechCrunch*. Retrieved March 25, 2018 from http://social.techcrunch.com/2013/11/20/googles-cerf-says-privacy-may-be-an-anomaly-historically-hes-right/

10. Financial Times Editorial Board. 2016. The taxman has strong grounds to test Uber's business model. *Financial Times*. Retrieved March 25, 2018 from https://www.ft.com/content/94738c64-c83d-11e6-8f29-9445cac8966f

11. David Ginsberg and Moira Burke. 2017. Hard Questions: Is Spending Time on Social Media Bad for Us? *Facebook Newsroom*. Retrieved from https://newsroom.fb.com/news/2017/12/hard-questions-is-spending-time-on-social-media-bad-for-us/

12. Glenn Beck. 2018. 1/18/18 – Tech Regs & Social Responsibility (w/ Moshe Vardi). *Glenn Beck*. Retrieved March 20, 2018 from http://www.glennbeck.com/content/audio/11818-tech-regs-social-responsibility-w-moshe-vardi/

13. Kevin Granville. 2018. Facebook and Cambridge Analytica: What You Need to Know as Fallout Widens. *The New York Times*. Retrieved from https://www.nytimes.com/2018/03/19/technology/facebook-cambridge-analytica-explained.html

14. Mike Hintze and Gary LaFever. 2017. *Meeting Upcoming GDPR Requirements While Maximizing the Full Value of Data Analytics*. Social Science Research Network, Rochester, NY. Retrieved March 26, 2018 from https://papers.ssrn.com/abstract=2927540

15. Michael Kassner. 2015. Anatomy of the Target data breach: Missed opportunities and lessons learned. *ZDNet*. Retrieved from http://www.zdnet.com/article/anatomy-of-the-target-data-breach-missed-opportunities-and-lessons-learned/

16. Katie Harbath. 2018. Hard Questions: Social Media and Democracy | Facebook Newsroom. Retrieved March 20, 2018 from https://newsroom.fb.com/news/2018/01/hard-questions-democracy/

17. Brendan I. Koerner. 2016. Inside the Cyberattack That Shocked the US Government. *WIRED Magazine*. Retrieved from https://www.wired.com/2016/10/inside-cyberattack-shocked-us-government/

18. Kotaro Hara, Abigail Adams, Kristy Milland, Saiph Savage, Chris Callison-Burch, and Jeffrey P. Bigham. 2018. A Data-Driven Analysis of Workers' Earnings on Amazon Mechanical Turk. In *ACM SIGCHI 2018*.

19. Yilun Lin, Johannes Schöning, and Brent Hecht. Examining "Death by GPS": A Systematic Analysis of Catastrophic Incidents Associated with Personal Navigation Technologies. In *Submitted to ACM SIGCHI 2017*.

20. Arvind Malhotra and Marshall Van Alstyne. 2014. The Dark Side of the Sharing Economy … and How to Lighten It. *Commun. ACM* 57, 11: 24–27. https://doi.org/10.1145/2668893

21. Farhad Manjoo. 2017. Facebook Conceded It Might Make You Feel Bad. Here's How to Interpret That. Retrieved March 26, 2018 from https://www.nytimes.com/2017/12/15/technology/facebook-blog-feel-bad.html

22. Sanika Mokashi, Svetlana Yarosh, and Gregory D. Abowd. 2013. Exploration of Videochat for Children with Autism. In *Proceedings of the 12th International Conference on Interaction Design and Children* (IDC '13), 320–323. https://doi.org/10.1145/2485760.2485839

23. National Highway and Transportation Safety Board. 2016. Distracted Driving. *NHTSA*. Retrieved March 21, 2018 from https://www.nhtsa.gov/risky-driving/distracted-driving

24. Cathy O'Neil. 2017. That free health tracker could cost you a lot. *Bloomberg*. Retrieved from http://www.charlotteobserver.com/opinion/op-ed/article137459473.html

25. Mike Orcutt. 2017. Blockchains Use Massive Amounts of Energy—But There's a Plan to Fix That. *MIT Technology Review*. Retrieved from https://www.technologyreview.com/s/609480/bitcoin-uses-massive-amounts-of-energybut-theres-a-plan-to-fix-it/

26. Radiolab. *Breaking News*. Retrieved March 21, 2018 from http://www.radiolab.org/story/breaking-news/

27. Robert Chesney and Danielle Citron. 2018. Deep Fakes: A Looming Crisis for National Security, Democracy and Privacy? *The Lawfare Blog*. Retrieved from https://www.lawfareblog.com/deep-fakes-looming-crisis-national-security-democracy-and-privacy

28. Adi Robertson. 2018. Reddit bans 'deepfakes' AI porn communities. *The Verge*. Retrieved from https://www.theverge.com/2018/2/7/16982046/reddit-deepfakes-ai-celebrity-face-swap-porn-community-ban

29. Matthew Rosenberg and Sheera Frenkel. 2018. Facebook's Role in Data Misuse Sets Off Storms on Two Continents. *The New York Times*. Retrieved March 20, 2018 from https://www.nytimes.com/2018/03/18/us/cambridge-analytica-facebook-privacy-data.html

30. Kate Starbird. 2017. Information Wars: A Window into the Alternative Media Ecosystem. *Medium*. Retrieved March 26, 2018 from https://medium.com/hci-design-at-uw/information-wars-a-window-into-the-alternative-media-ecosystem-a1347f32fd8f

31. Omer Tene. 2013. Vint Cerf is Wrong. Privacy Is Not An Anomaly. *Center for Internet and Society*. Retrieved from http://cyberlaw.stanford.edu/publications/vint-cerf-wrong-privacy-not-anomaly

32. Craig Timberg. 2017. Lawmakers demand answers about Strava 'heat map' revealing military sites. *The Washington Post: The Switch*. Retrieved from https://www.washingtonpost.com/news/the-switch/wp/2018/01/31/lawmakers-demand-answers-about-strava-heat-map-revealing-military-sites/?utm_term=.f8205678e39b

33. Sherry Turkle. 2017. Perspective | Why these friendly robots can't be good friends to our kids. *Washington Post*. Retrieved March 21, 2018 from https://www.washingtonpost.com/outlook/why-these-friendly-robots-cant-be-good-friends-to-our-kids/2017/12/07/bce1eaea-d54f-11e7-b62d-d9345ced896d_story.html

34. United Nations Development Programme. 2011. *Climate Change in Least Developed Countries*. United Nations. Retrieved from http://www.undp.org/content/dam/undp/library/corporate/fast-facts/english/FF-Climate-Change-in-Least-Developed-Countries.pdf

35. Moshe Y. Vardi. 2016. Are robots taking our jobs? *The Conversation*. Retrieved March 20, 2018 from http://theconversation.com/are-robots-taking-our-jobs-56537

36. Moshe Y. Vardi. The Moral Imperative of Artificial Intelligence. Retrieved March 20, 2018 from https://cacm.acm.org/magazines/2016/5/201608-the-moral-imperative-of-artificial-intelligence/fulltext

37. Tiffany C. Veinot, Jessica S. Ancker, Courtney Lyles, Andrea G. Parker, and Katie A. Siek. 2017. Good Intentions Are Not Enough: Health Informatics Interventions That Worsen Inequality. In *WISH @ AMIA 2017*.

38. Svetlana Yarosh, Anthony Tang, Sanika Mokashi, and Gregory D. Abowd. 2013. "Almost Touching": Parent-child Remote Communication Using the Sharetable System. In *Proceedings of the 2013 Conference on Computer Supported Cooperative Work* (CSCW '13), 181–192. https://doi.org/10.1145/2441776.2441798